\documentclass
[showpacs,superscriptaddress,prl,twocolumn,tightenlines,balancelastpage,10pt,a4paper]{revtex4}%
\usepackage{amsmath}
\usepackage{amsfonts}
\usepackage{amssymb}
\usepackage{graphicx}%
\begin{document}
\title{Experimental Quantum Secret Sharing and Third-Man Quantum Cryptography}
\author{Yu-Ao Chen}
\affiliation{Hefei National Laboratory for Physical Sciences at Microscale, Department of
Modern Physics, University of Science and Technology of China, Hefei, 230027,
People's Republic of China}
\affiliation{Physikalishes Institut, Universit\"{a}t Heidelberg, Philosophenweg 12, D-69120
Heidelberg, Germany}
\author{An-Ning Zhang}
\affiliation{Hefei National Laboratory for Physical Sciences at
Microscale, Department of Modern Physics, University of Science
and Technology of China, Hefei, 230027, People's Republic of
China}
\author{Zhi Zhao}
\affiliation{Hefei National Laboratory for Physical Sciences at
Microscale, Department of Modern Physics, University of Science
and Technology of China, Hefei, 230027, People's Republic of
China}
\affiliation{Physikalishes Institut, Universit\"{a}t
Heidelberg, Philosophenweg 12, D-69120 Heidelberg, Germany}
\author{Xiao-Qi Zhou}
\affiliation{Hefei National Laboratory for Physical Sciences at Microscale, Department of
Modern Physics, University of Science and Technology of China, Hefei, 230027,
People's Republic of China}
\author{Chao-Yang Lu}
\affiliation{Hefei National Laboratory for Physical Sciences at Microscale, Department of
Modern Physics, University of Science and Technology of China, Hefei, 230027,
People's Republic of China}
\author{Cheng-Zhi Peng}
\affiliation{Hefei National Laboratory for Physical Sciences at
Microscale, Department of Modern Physics, University of Science
and Technology of China, Hefei, 230027, People's Republic of
China}
\author{Tao Yang}
\affiliation{Hefei National Laboratory for Physical Sciences at
Microscale, Department of Modern Physics, University of Science
and Technology of China, Hefei, 230027, People's Republic of
China}
\author{Jian-Wei Pan}
\affiliation{Hefei National Laboratory for Physical Sciences at Microscale, Department of
Modern Physics, University of Science and Technology of China, Hefei, 230027,
People's Republic of China}
\affiliation{Physikalishes Institut, Universit\"{a}t Heidelberg, Philosophenweg 12, D-69120
Heidelberg, Germany}


\begin{abstract}
Quantum secret sharing (QSS) and third-man quantum cryptography
(TQC) are essential for advanced quantum communication, however,
the low intensity and fragility of multi-photon entanglement
source in previous experiments have made their realization an
extreme experimental challenge. Here, we develop and exploit a
ultra-stable high intensity source of four-photon entanglement to
report an experimental realization of QSS and TQC. The technology
developed in our experiment will be important for future
multi-party quantum communication.
\end{abstract}
\maketitle

With the development of technology, quantum cryptography
\cite{BB84,Ekert} could well be the first commercial application
of quantum communication \cite{gisinRMP}. To extend quantum
communication to more realms, advanced protocols are needed. Among
them, Quantum secret sharing (QSS) \cite{sharing1} and third-man
quantum cryptography (TQC) \cite{TQC} are two essential protocols.
QSS \cite{sharing1} is a protocol to split a message into several
parts so that no subset of parts is sufficient to read the
message, but the entire set is. In the scheme, three parties
Alice, Bob and Charlie first share a three-photon entangled state.
Charlie can then force Alice and Bob to cooperate to be able to
establish the secret key with him by performing proper
polarization measurements on his photon and announcing which
polarization basis he has chosen. In a similar manner, in TQC
\cite{TQC} the third-man, Charlie, can control whether Alice and
Bob can communicate in a secure way while he has no access
whatsoever on the content of the communication between Alice and
Bob.

Although QSS and TQC are essential for advanced quantum
communication, the low intensity and fragility of multi-photon
entanglement source in previous experiments
\cite{3GHZ,4pho,4phozhao} have made their realization an extreme
experimental challenge. While a variant of QSS, i.e. quantum state
sharing \cite{lo}, has been reported in both photonic and continue
variable systems \cite{zhao04,lam}, till now solely the principle
feasibility of an experimental realization of QSS using pseudo-GHZ
states was shown \cite{GisinQSS}. Here, developing and exploiting
an ultra-stable high intensity source of four-photon entanglement
we report the first experimental realization of the QSS
\cite{sharing1} and TQC \cite{TQC} schemes.

To see the necessity of QSS, suppose Alice and Bob are sent to
Beijing as two separate outstations by Charlie who is in the
parent company in Hefei. When Charlie wants to send a business
instruction to Beijing, the information should be encrypted
because it is a business secret. However, there is a risk that if
one of the two people received part of the information, they can
dishonestly sell it to other company for money. Remarkably, QSS
\cite{sharing1} can provide a novel way to solve this problem.

\begin{figure}
[ptb]
\begin{center}
\includegraphics[
width=2.0845in]%
{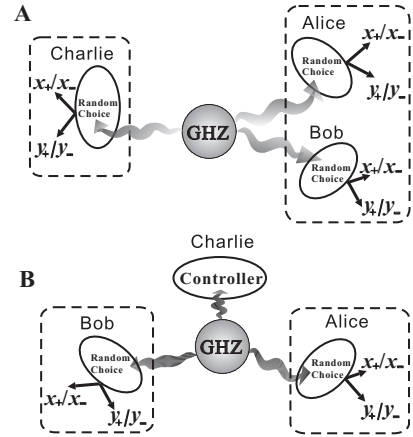}%
\caption{Schematic drawing of the QCC and TQC schemes.}%
\label{fig1}%
\end{center}
\end{figure}

The working principle of QSS is shown in Fig. 1A. Suppose that
Alice, Bob, and Charlie each holds one photon from a GHZ triplet
that is in the
state:%
\begin{equation}
|\Psi\rangle_{abc}=\frac{1}{\sqrt{2}}\left(  |H\rangle_{a}|H\rangle
_{b}|H\rangle_{c}+|V\rangle_{a}|V\rangle_{b}|V\rangle_{c}\right) \label{GHZ}%
\end{equation}%
where H (V) denotes horizontal (vertical) linear polarization.
Each of them randomly performs a projection measurement on their
own photons either along the linear polarization basis
$x_{+}/x_{-}$ rotated by 45$^{0}$ with respect to the original
basis, or along the circular polarization basis $y_{+}/y_{-}$
(right-hand/left-hand). These polarization basis can be expressed
in terms of
the original ones as%
\begin{align}
\left\vert x_{\pm}\right\rangle  &  =\frac{1}{\sqrt{2}}\left(  |H\rangle
\pm|V\rangle\right) \nonumber\\
\left\vert y_{\pm}\right\rangle  &  =\frac{1}{\sqrt{2}}\left(
|H\rangle\pm i|V\rangle\right) \label{basis}%
\end{align}%
For convenience we will refer to a measurement of linear
polarization $x_{+}/x_{-}\;$as an $x$ measurement and one of
circular polarization $y_{+}/y_{-}$ as a $y$\ measurement.

Representing the GHZ state (\ref{GHZ}) in the new states by using
\ref{basis}, we can obtain the perfect correlations among Alice,
Bob and Charlie in some certain combinations of measurement basis.
For example, in
an $xxx$ measurement the state (\ref{GHZ}) may be expressed as:%
\begin{align}
|\Psi\rangle_{abc}  & =\frac{1}{2}\left(  |x_{+}\rangle_{a}|x_{+}\rangle
_{b}+|x_{-}\rangle_{a}|x_{-}\rangle_{b}\right)  |x_{+}\rangle_{c}\nonumber\\
& +\left(  |x_{+}\rangle_{a}|x_{-}\rangle_{b}+|x_{-}\rangle_{a}|x_{+}%
\rangle_{b}\right)  |x_{-}\rangle_{c}\label{represent}%
\end{align}%
This expression implies, first, that any specific result obtained
in any two-photon joint measurement is maximally random. For
example, Alice's and
Charlie's photons will exhibit polarizations $x_{+}x_{-},x_{-}x_{+},x_{+}%
x_{+}$\ or $x_{-}x_{-}$\ with the same probability of 25\% as well as Bob's
and Charlie's photons. Second, given any two results of measurements on any
two photons, they can predict with certainty the result of the corresponding
measurement performed on the other one. For example suppose Alice's photon and
Bob's photon exhibit different polarizations. Then by the second term in
(\ref{represent}), Charlie's photon will definitely be\ $x_{-}%
$\ polarized. Therefore, Alice and Bob can exploit this perfect correlation to
jointly create the secret key with Charlie.

Let us now analyze the quantum correlations of the state \ref{GHZ}
for all the other combinations of measurement basis, i.e. for
$xyy$, $yxy$, $yyx$, $xxy$, $xyx$, $yxx\;$and $yyy$\ measurements.
In the same way, one can easily verify the following fact: While
in the last four combinations the results that Alice, Bob and
Charlie can get are completely independent (random) to each other,
similar perfect correlations also exist in the rest three
combinations (i.e. $xyy$, $yxy$ and $yyx$). Again, in these three
cases, if Alice and Bob know which measurement basis Charlie has
chosen (i.e. $x$ or $y$), they can cooperatively determine what
Charlie's result is.

In this way, if all of them randomly select the $x$, $y$
polarization basis to measure their own photons, then by
announcing publicly which basis they have chosen, and only keeping
those events with the right combinations of measurement basis,
i.e. the $xxx$, $xyy$, $yxy$ and $yyx$ combinations, Alice and Bob
can thus jointly establish the secret key with Charlie. In
details, we can use these measurement results to generate raw keys
by the following encoding rule: In the $xxx$ measurement, Charlie
encodes $x_{-}$($y_{+}$) as 1 and $x_{+}$($y_{-}$) as 0, while
Alice and Bob encode $x_{+}$($y_{+}$) as 1 and $x_{-}$($y_{-}$) as
0. For the other combinations, Alice, Bob and Charlie all encode
the result $x_{+}$($y_{+}$) as 1 and $x_{-}$($y_{-}$) as 0.

In the QSS scheme, its security can be guaranteed by randomly
choosing the measurement basis and testing the quantum bit error
rate (QBER) of the raw keys. Moreover, from the above
analysis it is clear that after Charlie uses the generated secret
key to encrypt a message, none of Alice and Bob can decrypt the
message with her/his individual key and it is only possible for
them to read out the encrypted message after performing a
cooperative Exclusive OR (XOR) operation.

Another important application of the above perfect correlations is
in the TQC \cite{TQC}. As in usual telecommunication, it is
expected that the necessary resource of single photons or quantum
entanglement in future realistic quantum communication will be
provided by some organizations such as a company or the
government. It is reasonable that the providers of quantum
resources would like to have some controls on the users. For
example, the providers would like to hold the right to control
whether the users can communicate in a secure way while, as a
regulation, they have no access whatsoever on the content of the
communication between the users. The TQC offers a satisfied way to
accomplish this purpose.

As it shown in Fig. 1B, in the TQC scheme Alice, Bob and Charlie
also need to share a three-photon GHZ state \ref{GHZ}. However,
Alice and Bob are now two users at two different locations, say,
Beijing and Shanghai, while Charlie in Hefei plays the role of a
provider. Similarly, each of them performs a projection
measurement on their own photons by randomly choosing either the
linear polarization basis $x_{+}/x_{-}$, or the circular
polarization basis $y_{+}/y_{-}$. From (\ref{represent}) we can
see if Charlie publicly announces his measurement results
including the basis chosen, Alice and Bob's photons can
immediately build up their own perfect correlation. Thus, Alice
and Bob can exploit this perfect correlation to create the secret
keys. This exactly corresponds to the entanglement-assisted BB84
protocol \cite{BB84}. Clearly, in this process Charlie has no idea
of the keys which Alice and Bob have created.

On the other hand, if Charlie does not want Alice and Bob to
generate the secret keys he could simply choose not to announce
his measurement results or not to make any measurement on his
photon. Then, if Alice and Bob continue to finish the $x$\ or $y$\
projection measurement, the results that they will get will be
completely random. Therefore, without the help of Charlie they
will fail to generate the secret key. Note that, if Alice and Bob
do not trust Charlie in the beginning, instead of performing a
$y$\ measurement on their own photons, they could perform a
measurement along the $H/V$ basis. In this way, they can manage to
generate some keys with perfect correlation, but they are insecure
-- indeed any eavesdropper can acquire the same key without being
detected. The above analysis shows that Charlie can successfully
control the generation of the secret keys between Alice and Bob
while has no access whatsoever on the content of the communication
between Alice and Bob.

\begin{figure}
[ptb]
\begin{center}
\includegraphics[
width=1.8in
]%
{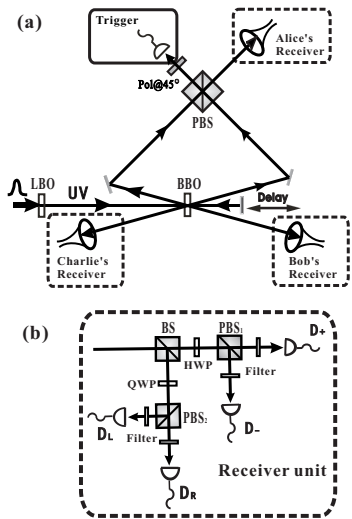}%
\caption{Experimental setup for QSS and TQC. (a) The UV pulse
generated by up-conversion inside the LBO crystal passes through
the BBO crystal twice to generate two pair of
polarization-entangled photons to prepare the desired three-photon
GHZ state. The UV laser with a central wavelength of 394 nm has a
pulse duration of 200fs, a repetition rate of 76 MHz, and an
average pump power of 430mw. By optimizing the collection
efficiency, we are able to observe about 2 fourfold coincidence
per second behind 3.6 nm filters (F) of central wavelength 788 nm.
(b) Each of Alice, Bob, and Charlie randomly performs a projection
measurement on her/his own photon. The BS is used to select basis
randomly to measure the photons. The HWP ($\lambda/2$) and QWP
($\lambda/4$) are used to perform $x_{+}/x_{-}$ and $y_{+}/y_{-}$
measurement. If and only if there is one click is registered at
each of the receivers and trigger, they individually record the measurement result.}%
\label{fig2}%
\end{center}
\end{figure}

The realization of QSS and TQC necessitates a ultra-stable high
intensity source of three-photon entanglement. The experimental
setup to generate three-photon entanglement is shown in Fig. 2(a).
An infrared pulse is focused properly on the LBO crystal
(LiB$_3$O$_5$) to achieve the best up-conversion efficiency
creating the pulse of ultraviolet (UV) light. Then the created UV
pulse passes through a beta-barium borate (BBO) crystal twice to
produce two polarization-entangled photon pairs, where both pairs
are in the state $\left\vert \Psi\right\rangle =1/\sqrt{2}\left(
|H\rangle|H\rangle +|V\rangle|V\rangle\right) $. One photon out of
each pair is then steered to a polarization beam splitter (PBS)
where the path lengths of each photon have been adjusted (by
scanning the Delay position) so that they arrive simultaneously.
After the two photons pass through the PBS, and exit it by a
different output port each, and there is no way whatsoever to
distinguish from which emission which of the photons originated,
then correlations due to four-photon GHZ entanglement $\left\vert
\Psi ^{4}\right\rangle =1/\sqrt{2}\left(
|H\rangle|H\rangle|H\rangle
|H\rangle+|V\rangle|V\rangle|V\rangle|V\rangle\right)  $ can be
observed \cite{4pho,4phozhao}. In the experiment, by performing a
$x_{+}$ polarization projective measurement onto one of the four
outputs, the remaining three photons are prepared in the desired
GHZ-state (\ref{GHZ}) with a visibility of 83\%.

To achieve the necessary ultra stability and high intensity,
various efforts have been made. Different from previous
four-photon experiments \cite{4pho,4phozhao}, to avoid the damage
to the up-conversion LBO crystal caused by the focusing laser
beam, we assemble the LBO crystal in a closed but transparent tube
of oxygen. Moreover, by using thicker and more rigid fiber holder
in a compact set-up and by focusing the UV pump onto the BBO
crystal, we achieve both better collection efficiency and
production rate of entangled photon pairs with a ultra-high
stability. In our experiment, after achieving the perfect time
overlap between the two photons coming into the PBS the
four-photon entanglement source with a 2 four-fold coincidence per
second can be stabilized for a couple of weeks. Thus, we managed
to finish all the required measurements without the need of
scanning the Delay position.

In order to realize the random choice of the measurement basis, we
let the photon pass through a 50-50 beam splitter (BS) as in Fig.
2(b). The half-wave plates (HWP) in front of PBS$_{1}$ is oriented
at 22.5$^{0}$ to measure the photon along the linear polarization
basis $x_{+}/x_{-}$, and the quarter-wave plate (QWP) in front of
PBS$_{2}$ is oriented at 45$^{0}$ to measure the photon along the
circular polarization basis $y_{+}/y_{-}$. In our experimental
verification of the QSS and TQC schemes, Alice, Bob and Charlie
only individually record the measurement results (including the
basis chosen) for those events where one and only one click is
registered at each of the four detectors.  All together, 13
single-photon detectors have been used during the whole
experiment.

In the QSS scheme, after the measurement run Alice, Bob and
Charlie announce the basis of their measurement results in public.
By only keeping those coincident events corresponding to an $xxx$,
$xyy$, $yxy$ or $yyx$ measurement, which occur in half of the
cases, they can generate the raw keys using the encoding rule as
discussed in this paper. In the experiment, Alice, Bob and Charlie
collected 327 579 bits of key each at a rate of a quarter bit/s.
To test the security, 10\% of the raw keys are used to calculate
the QBER by checking if each triplet is coincident by $Alice\oplus
Bob=Charlie$\ in which $\oplus$\ plays as the operator of XOR. The
QBER is observed to be 12.9\%, which is sufficient to ensure the
security of the QSS scheme \cite{gisinRMP}.

For correcting the remaining errors while maintaining the secrecy
of the keys, various classical error correction and privacy
amplification schemes can be used. We implemented a simple error
reduction scheme requiring only little communication between
Alice, Bob and Charlie \cite{BB92}. Each of them arranges their
keys in blocks of bits and evaluates the bit parity of the blocks
(a single bit indicating an odd or even number of ones in the
block). The parities are compared by $Alice\oplus Bob=Charlie$\ in
public, and the blocks with agreeing parities are kept after
discarding one bit per block. Since parity checks reveal only odd
occurrences of bit errors, a fraction of errors remains. The
optimal block length $n$ should be determined by a compromise
between key losses and remaining bit errors. In order to obtain a
final key with a low QBER, we implemented the parity check for two
times and the first we use the block length $n$ of 2. After that
we got 117 616 bits of key with a QBER of 2.2\%. Again by the
block length $n$ of 8, we got 87 666 bits of key with a QBER of
0.35\%.

Finally, Charlie uses the corrected key to transmit a 76 160-bit
large image to Alice and Bob via two one-time-pad protocols,
utilizing a bitwise XOR combination of message and key data. As
shown in Fig. 3, Alice and Bob can't get any information only with
their own keys. But they can read it with few
errors by cooperating.%

\begin{figure}
[ptb]
\begin{center}
\includegraphics[
width=3.3in
]%
{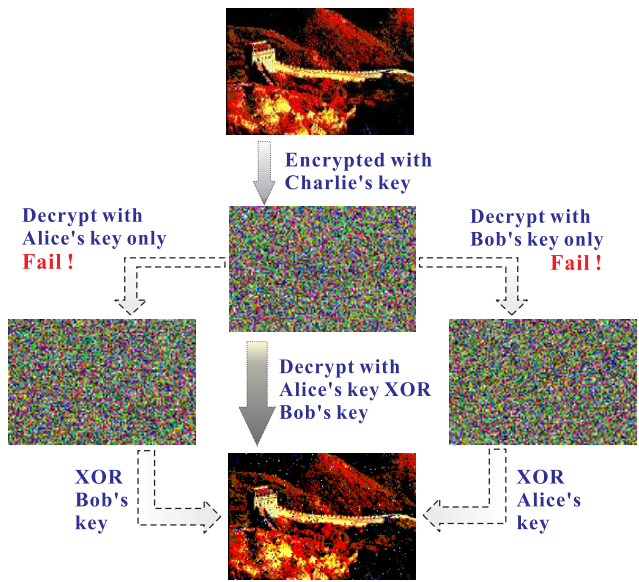}%
\caption{Sharing a secret image. Charlie encrypts the image of the
``Great Wall'' via bitwise XOR operation with his key and
transmits the encrypted image to Alice and Bob via the computer
network. None of Alice and Bob can decrypt the image with her or
his own key, but they can cooperate to decrypt the image. The
final image cooperatively obtained by Alice and Bob shows only a
few errors due to the remaining bit errors in
the keys (0.35\%).}%
\label{fig4}%
\end{center}
\end{figure}

We now show how the same experimental data can also be used to
provide an experimental demonstration of TQC. In the TQC, if
Charlie allows Alice and Bob to generate the secret keys with each
other, he will faithfully announce all his information, not only
the basis he has chosen but also his measurement results. Knowing
Charlie's measurement results, Alice then performs an XOR
operation between her and Charlie's keys. Thus, using the same
experimental data Alice and Bob can each obtain 327 579 bits of
raw keys with a QBER of 12.9\%. After QBER checking and error
reduction, they finally obtained 87 666 bits cured keys with a
QBER of 0.35\%. And, if Charlie does not want Alice and Bob to
generate the secret key, he simply chooses not to announce his
measurement results, or do not perform any measurement on his
photon. Without knowing Charlie's results, the only thing Alice
and Bob can do is to randomly guess Charlie's results and continue
the same encoding and error reduction procedure. In our
experiment, after performing twice error reductions, the QBER
remains 49.999\%. All these together clearly confirm that Charlie
can successfully control the secure communication between Alice
and Bob.

We thus for the first time experimentally demonstrated the QSS and
TQC. Compared to the quantum cryptography based on single photons,
the QSS and TQC schemes allow richer and more flexible quantum
communications. First, in the multi-party QSS any individual can
force the others to cooperate to be able to establish the secret
keys with her/him \cite{sharing1}. Moreover, with the future
development of quantum repeaters \cite{repeater} one can achieve
the QSS or TQC over large distances, without worrying about the
effects of attenuation and noise on, say, single photons sent
through a long optical fiber. Finally, the entanglement-assisted
QSS and TQC have the advantage of still being possible in
situations where, for example three parties, Alice, Bob and
Charlie, after sharing their storable multi-particle entanglement
\cite{DLCZ,Kuzmich}, have wandered about independently and no
longer know each others' locations. They cannot reliably send
single particles to each other, if they do not know where others
are; but they can still realize QSS and TQC, by broadcasting the
classical information to all places where they might be.
Therefore, while in large scale realization further practical
investigations are still necessary for improving the limited
multi-fold coincidence rate, we believe that the QSS and TQC could
very well be tomorrow's technology for advanced quantum
communication.

This work was supported by the NNSF of China, the CAS, the
Alexander von Humboldt Foundation, the Marie Curie Excellence
Grant of the EU and the Deutsche Telekom Stiftung.

\end{document}